\documentclass[reprint,superscriptaddress,showpacs,amsmath,amssymb,aps,floatfix,nolongbibliography]{revtex4-2}
\usepackage[utf8]{inputenc}
\usepackage[T1]{fontenc}
\usepackage[english]{babel}
\usepackage{amsmath}
\usepackage{graphicx}		
\usepackage{natbib}
\usepackage{textcomp}
\usepackage{gensymb}
\usepackage[hidelinks]{hyperref}
\usepackage{xcolor}
\usepackage{siunitx}
\usepackage{mathrsfs}
\usepackage{multirow}
\usepackage{amsthm}
\usepackage{float}
\usepackage{xspace}
\usepackage[normalem]{ulem}
\theoremstyle{definition}

\newcommand{\Emph}[1]{\textbf{#1}}
\newcommand*{\supprime}{\textsuperscript{\everymodeprime}\xspace}
\newcommand*{\ssupprime}{\textsuperscript{\everymodeprime\everymodeprime\xspace}}
\graphicspath{{figs-perspective/}}
\newcommand*{\everymodeprime}{\ensuremath{\prime}} 

\begin{document}
% \title{A unified framework of metastability in neuroscience}
%\title{A dynamics-based perspective on metastability in neuroscience}
\title{Dynamical properties and mechanisms of metastability: a perspective in neuroscience}
\author{Kalel L. Rossi}
\affiliation{Theoretical Physics/Complex Systems, ICBM, Carl von Ossietzky University Oldenburg, Oldenburg, Lower Saxony, Germany}
\author{Roberto C. Budzinski}
\affiliation{Department of Mathematics, Western University, London, Ontario, Canada}
\affiliation{Western Institute for Neuroscience, Western University, London, Ontario, Canada}
\affiliation{Western Academy for Advanced Research, Western University, London, Ontario, Canada}
\author{Everton S. Medeiros}
\affiliation{Theoretical Physics/Complex Systems, ICBM, Carl von Ossietzky University Oldenburg, Oldenburg, Lower Saxony, Germany}
\author{Bruno R. R. Boaretto}
\affiliation{Institute of Science and Technology, Federal University of São Paulo, São José dos Campos, São Paulo, Brazil}
\author{Lyle Muller}
\affiliation{Department of Mathematics, Western University, London, Ontario, Canada}
\affiliation{Western Institute for Neuroscience, Western University, London, Ontario, Canada}
\affiliation{Western Academy for Advanced Research, Western University, London, Ontario, Canada}
\author{Ulrike Feudel}
\affiliation{Theoretical Physics/Complex Systems, ICBM, Carl von Ossietzky University Oldenburg, Oldenburg, Lower Saxony, Germany}

\begin{abstract}
Metastability, characterized by a variability of regimes in time, is a ubiquitous type of neural dynamics. It has been formulated in many different ways in the neuroscience literature, however, which may cause some confusion. In this Perspective, we discuss metastability from the point of view of dynamical systems theory. We extract from the literature a very simple but general \textit{definition} through the concept of \textit{metastable regimes} as long-lived but transient epochs of activity with unique dynamical properties. This definition serves as an umbrella term that encompasses formulations from other works, and readily connects to concepts from dynamical systems theory. This allows us to examine general dynamical properties of metastable regimes, propose in a didactic manner several \textit{dynamics-based mechanisms} that generate them, and discuss a theoretical tool to characterize them quantitatively. This perspective leads to insights that help to address issues debated in the literature and also suggest pathways for future research.
\end{abstract} 

\maketitle

% \section{Metastability is important, but what is it actually?}
% \section{Establishing a framework for metastability}
\section{Introduction}

Time-series of neural activity often reveal series of transitions between experimentally observable regimes with unique dynamical properties. For example, as subjects fall asleep, their brains progress through a series of well-defined patterns, from the 11-15 Hz sleep spindle \cite{contreras1996control}, to the large low-frequency rhythms of deep sleep \cite{steriade1993thalamocortical}, and to waking-like activity during rapid eye movement (REM) sleep \cite{jouvet1979does}. These sleep stages, or regimes, can also occur on shorter timescales, as subjects transition from passive rest to active perception \cite{poulet2008internal}. This ubiquitous phenomenon of regime switching has prompted connections to the concept of \textit{metastability} in physics and dynamical systems theory. Many works have demonstrated metastability in neural systems across different spatiotemporal scales and species \cite{michel2017eeg, vandeville2010eeg, lehmann1987eeg, jones2007natural, lacamera2019cortical, mazzucato2019expectation, recanatesi2021metastable, brinkman2022metastable, abeles1995cortical, seidemann1996simultaneously, jercog2017updown, luczak2007sequential, mazor2005transient, sasaki2007metastability, mashour2020conscious, dehaene2005ongoing, hudson2014recovery, tognoli2014metastable, popa2009constracting, curtis2015initiation, fernandez2020sleep, caruso2023single, lang2023temporal}.

Understanding the mechanisms based on dynamical systems theory that can generate such metastable regimes is crucial, especially as this knowledge can aid in the development of techniques to predict transitions between regimes and to possibly control them. The literature in neuroscience has made crucial advancements in this regard \cite{tognoli2014metastable, fingelkurts2017information, graben2019metastable, cavanna2018dynamic, brinkman2022metastable, fonollosa2015learning, deco2016metastability}, but differences in formulation between works can lead to some confusion. For instance, the explicit definition of metastability is not totally clear in the neuroscience literature (see Sec.~\ref{sec:currentformulations} for more). Further, discussions on the dynamics-based mechanisms for metastability are usually restricted to only a few possible mechanisms, depending on the context and formulation of each work.  

In this Perspective, we aim to address these issues by discussing the neuroscience literature through the lens of dynamical systems theory. First, we extract from the literature a very simple but general definition of metastability, based on \textit{metastable regimes}. Regimes are epochs of a time-series identified in each work that have unique dynamical properties. They are considered metastable when they are \textit{long-lived but transient}. This idea is widespread among works not only in neuroscience, but is also similar to the well-defined view in physics and dynamical systems theory \cite{olivieri2005large, callen1991thermodynamics, yorke1979metastable}. The characterization as long-lived is specific to the system being studied, and discussed in details later. We argue that all metastable regimes share a crucial defining property in state space, which is the space spanned by all variables of a system (explained in detail in Sec.~\ref{sec:stability_and_invariance}). Simply put, metastable \textit{regimes} correspond to metastable \textit{regions} in state space, in which trajectories have a high probability of remaining inside. This idea can be formulated theoretically through the concept of almost-invariant regions \cite{froyland2005statistically, dellnitz2003congestion} (see Sec.~\ref{sec:framework} for details). 
% In this Perspective, we aim to address these issues by proposing a \textit{coherent framework for metastable dynamics} that unifies previous formulations and mechanisms from both the neuroscience and dynamical systems literatures. First, we highlight the idea of \textit{metastable regimes} that is common to observations and formulations of metastability in neuroscience. Regimes are epochs of a time-series identified in each work that have unique dynamical properties, and they are considered metastable when they are \textit{long-lived but transient}. This idea is widespread among works not only in neuroscience, but is also similar to the well-defined view in physics and dynamical systems theory \cite{olivieri2005large, callen1991thermodynamics, yorke1979metastable}. The characterization as long-lived is specific to the system being studied, and the framework we propose is flexible to this decision. We argue that all metastable regimes share a crucial defining property in state space, which is the space spanned by all variables of a system (explained in detail in Sec.~\ref{sec:stability_and_invariance}). Simply put, metastable \textit{regimes} correspond to metastable \textit{regions} in state space, in which trajectories have a high probability of remaining inside. This idea can be formulated theoretically through the concept of almost-invariant regions \cite{froyland2005statistically, dellnitz2003congestion} (see Sec.~\ref{sec:framework} for details). 

Associating metastable regimes in time to metastable regions in state space, which in turn are defined as almost-invariant \cite{froyland2005statistically, dellnitz2003congestion}, allows for a single coherent view on metastability. It also allows for a direct connection to well-known dynamics-based mechanisms from dynamical systems theory. Here, we discuss and compare several of these mechanisms in a didactic manner. By looking into them together through a general definition, we are able to add important insights into issues debated in the literature and propose new research directions to generate experimentally testable and falsifiable hypotheses about metastability in the brain, potentially including phenomena such as sleep, seizures, and computations in neural circuits.

\section{Short summary of metastability in the brain}
\subsection{Observations of metastability} \label{sec:observations}
We now review key observations that have helped to establish metastability as an important dynamical phenomenon. We focus on examples of metastability characterized by the switching between long-lived regimes of activity. These regimes are epochs of observations with unique dynamical properties identified by the authors of each work, and their characterization as long-lived depends on the specific application being considered. 
Further, as a note of terminology, these regimes are often called states in the neuroscience literature, but we avoid this here because it conflicts with already well-established terminologies in other fields of science. Instead, we believe regime is equally descriptive, and less ambiguous. In this section, we only maintain state as a synonym for regime when it refers to a well-established name, such as EEG microstates. 

\begin{figure*}[hbt]
    \centering
    \includegraphics[width=\textwidth]{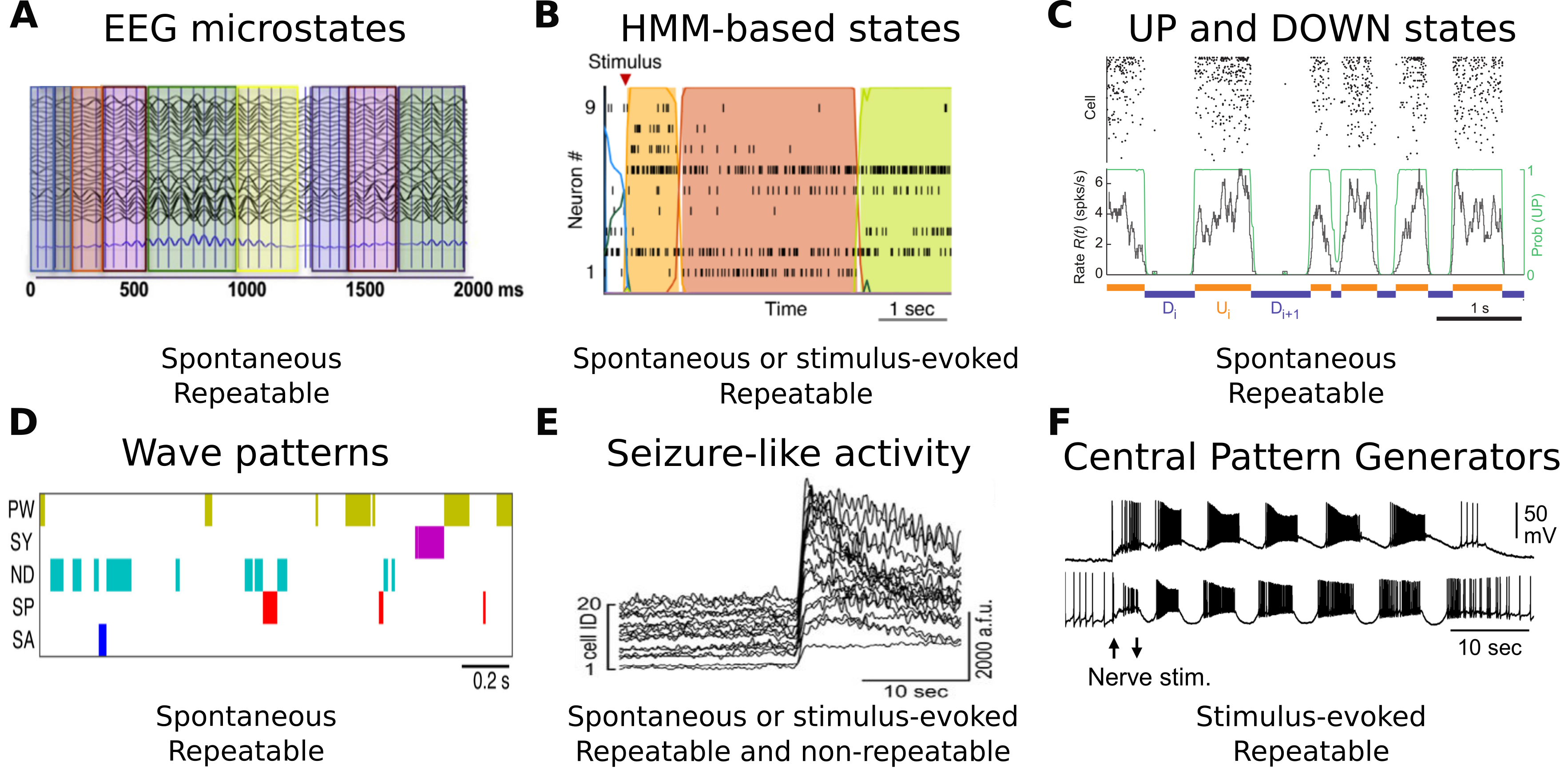}
    \caption{\textbf{Brain activity typically evolves as a sequence of well-defined regimes that are transient but long lived.} The panels illustrate observations taken from the literature, namely EEG microstates \cite{michel2017eeg} (A), firing rate states identified via hidden-Markov-modelling (HMM) \cite{brinkman2022metastable} (B), UP and DOWN states \cite{jercog2017updown} (C), cortical wave patterns \cite{townsend2015emergence} (D), seizure-like activity \cite{wenzel2019acute} (E), and bursting in central pattern generators \cite{sakurai2016recruitment} (F). Each observation is explained in detail in the main text, and is classified into subtypes of metastability that we define in the Sec.~\ref{sec:subtypes}. }
    \label{fig:observationsmetastability}
\end{figure*}

Figure \ref{fig:observationsmetastability} illustrates some observations showing time-series with metastable regimes, along with two corresponding characteristics: (i) whether the transitions between the regimes occur during resting conditions (spontaneously) or if they are evoked by some stimulus; and (ii) whether the metastable regimes occur more than once in the observation (repeat) or not.  

As a first example, Fig.~\ref{fig:observationsmetastability}A shows time-series of electroencephalography (EEG) measurements performed in resting humans with eyes closed (taken from Ref.~\cite{michel2017eeg}). Colors indicate \Emph{EEG microstates}, identified from the spatial configuration of the electric potential amplitude of each electrode. These microstates remain almost constant for roughly $\SI{100}{\milli\second}$ \cite{vandeville2010eeg, lehmann1987eeg}, and then give way to other microstates.

Figures \ref{fig:observationsmetastability}B-C show results based on firing rates of neurons. In particular, Fig.~\ref{fig:observationsmetastability}B shows exemplary sequences of regimes adapted from Ref.~\cite{brinkman2022metastable}. The regimes are characterized by roughly stationary behavior in the firing rates of neurons in the gustatory cortex of rats, and are identified via the technique of \Emph{hidden Markov modelling} (HMM). Each regime lasts for roughly an order of magnitude longer than the transitions between any regime \cite{jones2007natural, lacamera2019cortical, mazzucato2019expectation, recanatesi2021metastable, brinkman2022metastable}. This is observed during both spontaneous and stimulus-evoked activity, and the sequence of such regimes is shown to encode the stimuli presented to the animals \cite{lacamera2019cortical, mazzucato2019expectation}. These regimes are proposed to serve as a ``substrate for internal computations" in the brain \cite{lacamera2019cortical}. Similar results have also been reported in the frontal cortex of monkeys during a delayed localization task \cite{abeles1995cortical, seidemann1996simultaneously}.

Figure~\ref{fig:observationsmetastability}C shows sequences of sustained firing (significant activity, \Emph{UP states}) and silence (\Emph{DOWN states}) in the firing rates of the deep layers of the somatosensory cortex of urethane-anesthetized rats (adapted from Ref.~\cite{jercog2017updown}). Each regime lasts for a significant time, with quick transitions between them \cite{jercog2017updown}. These regimes are ubiquitously observed in spontaneous activity \cite{luczak2007sequential, jercog2017updown}. 

Figure \ref{fig:observationsmetastability}D shows a sequence of complex \Emph{wave patterns} identified in the cerebral cortex of anesthetized marmoset monkeys, taken from Ref. \cite{townsend2015emergence}. Starting from local field potential data from multiple electrodes, the authors created a spatial map for the phases of the oscillations, filtered at the $\delta$ band ($1-\SI{4}{\hertz}$), from which they identified different spatiotemporal patterns, classified as plane waves (PW), synchrony (SY), node (ND), spiral (SP), and saddle (SA) \cite{townsend2015emergence}. These patterns repeat across the time series, but the probability of switching from one pattern to another differs from pattern to pattern, i.e., there is preferential switching between patterns \cite{townsend2015emergence}.
% The firing rate of neurons in slice cultures of the hippocampal CA3 region was found to evolve also as sequences of discrete, long-lasting states, separated by quick transitions \cite{sasaki2007metastability} (Fig. \ref{fig:observationsmetastability}D). The figure was taken from \cite{sasaki2007metastability} (Copyright 2007 Society for Neuroscience). Each state is identified by clustering algorithms applied to the space spanned by the principal components obtained from Principal Component Analysis (\Emph{PCA}) of the firing rates. The figure shows the raster plot of the firing, and the total activity, with colors denoting each state \cite{sasaki2007metastability}.

Mazor and Laurent \cite{mazor2005transient} (not included in the figure) measured the firing rate of neurons in the antennal lobe of locusts as the animals were presented with a variety of odor pulses. Their results suggest that, for long pulses (lasting more than $\SI{2}{\second}$), spiking activity in the antennal lobe switches between a baseline fixed point and an odor-specific fixed-point, which can last for a few seconds. Interestingly, the transition epochs between the two fixed points were found to contain the most amount of information about the stimuli, suggesting an important role for them in neural computations \cite{mazor2005transient, rabinovich2008transientdynamics}. This suggests a functional role of transitions between metastable regimes for information processing \cite{mazor2005transient, ashwin2005when}.

Figure \ref{fig:observationsmetastability}E shows the neural activity recorded by two-photon calcium imaging in mouse neocortex \cite{wenzel2019acute} (Copyright 2019 Society for Neuroscience). The neural population is being invaded by a propagating \Emph{seizure-like activity}, which appears in the middle of the time-series, and is characterized by sustained firing of a large number of neurons. Traveling as a wave, it transiently replaces the baseline regime, in which firing is more sparse and distributed. In this study, a pharmacological agent was applied to a specific region of the cortex to induce the seizure-like activity.

% Figure \ref{fig:observationsmetastability}E shows that metastable states also occur in the context of consciousness. The global neuronal workspace theory of consciousness proposes that we become conscious of a certain object (e.g. a stimulus) when the representation of that object is broadcast from local processing regions into a variety of spatially distributed regions, which form the global workspace \cite{mashour2020conscious}. This global broadcast is known as \Emph{ignition}, and is achieved through the sustained firing of the involved areas; this is shown in panel E for the local-field potential (LFP) and spikes in area D2 of the model studied in \cite{dehaene2005ongoing}. The figure is adapted from \cite{dehaene2005ongoing}. The activity of workspace neurons is thus characterized by discrete episodes of spontaneous coherent activation, with sustained firing, and quick transitions between them \cite{michel2017eeg}.

Figure \ref{fig:observationsmetastability}F shows the swim motor pattern in intracellular recordings of two neurons that belong to the swim central pattern generator of the mollusk \textit{Tritonia} \cite{sakurai2016recruitment} (adapted from Ref. \cite{sakurai2016recruitment}). These neurons respond by \Emph{bursting} when nerve PdN3 is stimulated. These bursts are phenomenologically similar to those shown later in Fig. \ref{fig:mechanismsmetastability}E (though their mechanisms might differ), and demonstrate the alternation between periods of sustained firing and periods of silence. The dynamics of central pattern generators often include metastable regimes \cite{marder2001central}. 

Another work (not included in the figure) has also found that the power spectrum of local-field potential (LFP) recordings in rats progresses as a sequence of relatively stationary regimes lasting for some time before rapidly transitioning to other regimes \cite{hudson2014recovery} (see e.g. Fig. 2 of Ref. \cite{hudson2014recovery}). This was observed as the rats recovered consciousness, when the concentration of an anesthetic was progressively decreased. The authors argue that the existence of well-defined metastable regimes is crucial for the fast recovery of consciousness \cite{hudson2014recovery}.

Additionally, we remark that several other regimes could be mentioned in this section, such as other types of seizures \cite{curtis2015initiation, babloyantz1986low}; sleep spindles \cite{fernandez2020sleep}, transient patterns of circular waves that repeatedly travel across the cortex during sleep and may aid the consolidation of memories \cite{muller2016rotating}, and the phases of local-field potentials in cats at rest \cite{tognoli2014metastable, popa2009constracting}. For further reviews on this topic, we refer the reader to Refs. \cite{tognoli2014metastable, brinkman2022metastable, cavanna2018dynamic, tsuda2009hypotheses}.

% We also note that some works mentioned previously may have not necessarily used the term metastability in their respective papers. \lm{Finally,} we note that some works interpret metastability differently than the idea of long-lasting states, and they were not included here.

\subsection{Current formulations of metastability}\label{sec:currentformulations}

\begin{figure*}[htb]
    \centering
    \includegraphics[width=\textwidth]{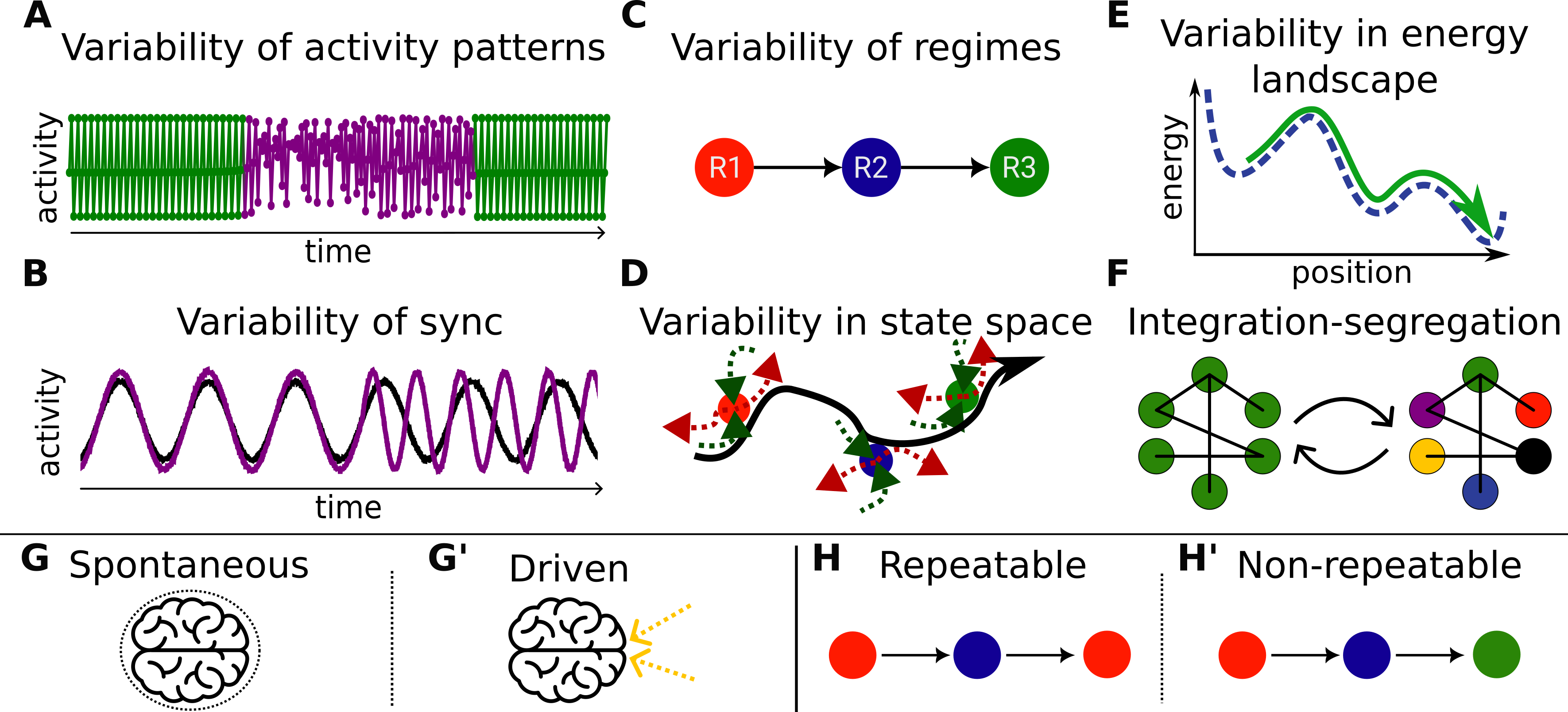}
    \caption{\textbf{Formulations of metastability in the neuroscience literature.} A common theme among these is the presence of transitions between certain aspects of the system's dynamics (e.g., between activity patterns). The upper part of the figure (A-F) illustrates what these aspects are in each formulation, and the bottom part (G-H\supprime) shows characteristics of the transitions. Further details are available in the main text.}
    \label{fig:definitionsmetastability}
\end{figure*}

% Several theoretical formulations have been discussed as a result of the observations of metastability in the brain. However, despite their large applicability, such formulations often depend on the specific context of their observations. To illustrate this issue, we depict in Figs. \ref{fig:definitionsmetastability}A-F the main formulations of metastability that we have identified in the neuroscience literature.
Several theoretical formulations have been employed to study the observations of metastability in the brain, but they have remained focused mostly on the specific context of their observations, and a general view is currently lacking. To illustrate this issue, we depict in Figs. \ref{fig:definitionsmetastability}A-F the main formulations of metastability that we have identified in the neuroscience literature.

We can generalize the formulations in Figs.~\ref{fig:definitionsmetastability}A-E as considering metastability to be a behavior with a type of variability in the dynamics. This variability can be directly observed as different \textit{patterns of activity} (Fig. \ref{fig:definitionsmetastability}A) \cite{friston1997transients, friston2000transients, varela2001brainweb, roberts2019metastable}; or \textit{patterns of synchronization} (Fig. \ref{fig:definitionsmetastability}B, where activity shifts between in-phase and out-of-phase) \cite{cabral2011role, deco2017dynamics, deco2016metastability, poncealvarez2015restingstate, aguilera2016extended}. As mentioned in the previous section, abstract regimes can be identified through techniques such as Hidden-Markov model \cite{mazzucato2015dynamics, lacamera2019cortical, brinkman2022metastable} or Principal Component Analysis (PCA) \cite{sasaki2007metastability}. Then, metastability is simply said to be the variability of these regimes (Fig. \ref{fig:definitionsmetastability}C, with each circle representing one distinct regime) \cite{mazzucato2015dynamics, lacamera2019cortical, afraimovich2010longrange, alderson2020metastable, lee2017linking, vasa2015effects, hellyer2014control, naik2017metastability, rabinovich2008transientcognitive, cavanna2018dynamic, werner2007metastability, bhowmik2013metastability}. In other works, metastability is said to occur due to variability of trajectories along different \textit{regions of state space} \cite{hudson2017metastability, graben2019metastable}; see Fig. \ref{fig:definitionsmetastability}D, where the black curve represents a trajectory passing through equilibria in colored circles, with their stable and unstable manifolds in dark green and red, respectively. For a detailed explanation of these concepts, we refer the reader to Sec.~\ref{sec:stability_and_invariance}. Metastability has also been defined as a variability along different positions on an \textit{energy landscape} (Fig. \ref{fig:definitionsmetastability}E) \cite{gili2018metastable, cavanna2018dynamic, aguilera2016extended, hudson2017metastability}. Another, more distinct, formulation of metastability is that of a behavior with \textit{integration-segregation}, in which neural assemblies transiently synchronize and desynchronize (Fig. \ref{fig:definitionsmetastability}F, where the nodes represent assemblies whose anatomical or functional connections correspond to lines, and their activity to colors, which alternate between synchronized and desynchronized) \cite{deco2015rethinking, fingelkurts2001operational, tognoli2014metastable, tognoli2014enlarging, bressler2016coordination, kelso2012multistability, hellyer2015cognitive}. 

Furthermore, we remark that, while some works explicitly mention a time-scale separation in metastability \cite{hudson2017metastability, brinkman2022metastable}, others do not. 
Works also differ in regard to the possible mechanisms giving rise to metastability. Some consider that the variability in the dynamics needs to be spontaneous, occurring in autonomous systems (Fig. \ref{fig:definitionsmetastability}G) \cite{sasaki2007metastability, kelso2012multistability, roberts2019metastable, fingelkurts2008brainmind, cavanna2018dynamic}, while others argue that the variability must be externally induced, occurring in non-autonomous systems (Fig. \ref{fig:definitionsmetastability}G\supprime) \cite{jercog2017updown, hudson2017metastability}. A combination of both is also found in Refs. \cite{brinkman2022metastable, friston2000transients, lacamera2019cortical}.  
Finally, another source of difference is in the repeatability of metastable regimes: some works consider metastable regimes only when they are repeatable  (Fig. \ref{fig:definitionsmetastability}H) \cite{graben2019metastable}, while others also consider non-repeatable (Fig \ref{fig:definitionsmetastability}H\supprime) regimes \cite{brinkman2022metastable}. Examples of these types of metastability are shown in the observations in Fig. \ref{fig:observationsmetastability}.

\subsection{The common thread} \label{sec:common-thread}

As we have seen in the previous sections, the explicit formulation of metastability can vary considerably depending on context. However, there is a common thread across all the observations and formulations. It starts with the notion of regimes as epochs of a time-series with unique dynamical properties. For instance, unique spatial configurations of the EEG electric field define unique EEG microstates \cite{michel2017eeg}, and UP states have an above-threshold firing rate, while DOWN states do not \cite{jercog2017updown}. 
% The properties defining the regimes are often often stationary or almost-stationary (i.e., having constant or almost-constant statistical moments).
Further, these regimes often have constant, or almost-constant properties throughout their duration.

% Then,  even if only qualitatively, . These timescales can be, for instance, a characteristic period of the EEG oscillations, or an average firing rate of neurons. As a consequence of this long duration, the regimes are called metastable.
Then, regimes are characterized as long-lived, even if only qualitatively, establishing thus the common thread across works: \Emph{metastable regimes are long-lived but transient}. Although sometimes implicit, we believe this common thread is the fundamental aspect of metastability as a dynamical behavior. This is also the view that has been held generally in physics \cite{olivieri2005large, callen1991thermodynamics}.

In this Perspective, we look into dynamics-based mechanisms that generate such metastable regimes, connecting these ideas with dynamical systems theory. To achieve this, we introduce some crucial concepts from this theory in the next section. 

% The dynamics are specified, both experimentally and theoretically, by observables (e.g., local-field potential in EEG, or neuronal firing rates). Properties can then be identified from these observables, such as wave patterns, hidden Markov states, or synchronization patterns, and they are often stationary or quasistationary in the statistical sense for the state's duration \cite{michel2017eeg, brinkman2022metastable, mazor2005transient} (as discussed in Sec. \ref{sec:observations}). 

\section{Metastability in state space}\label{sec:nonlinear}

\subsection{Stability and invariance - initial concepts}\label{sec:stability_and_invariance}

A dynamical system is a set of $N$ variables together with the rules that dictate their time evolution. These variables may be observables identified from experiments, for instance. At each instant, the values of these $N$ variables define the \textit{state} $\mathbf{s}$ of the system. The $N$-dimensional abstract space containing all possible states of a system is called the \textit{state space}. In this space, each dimension corresponds to a variable of the system. 

\begin{figure*}[htb]
    \centering
    \includegraphics[width=\textwidth]{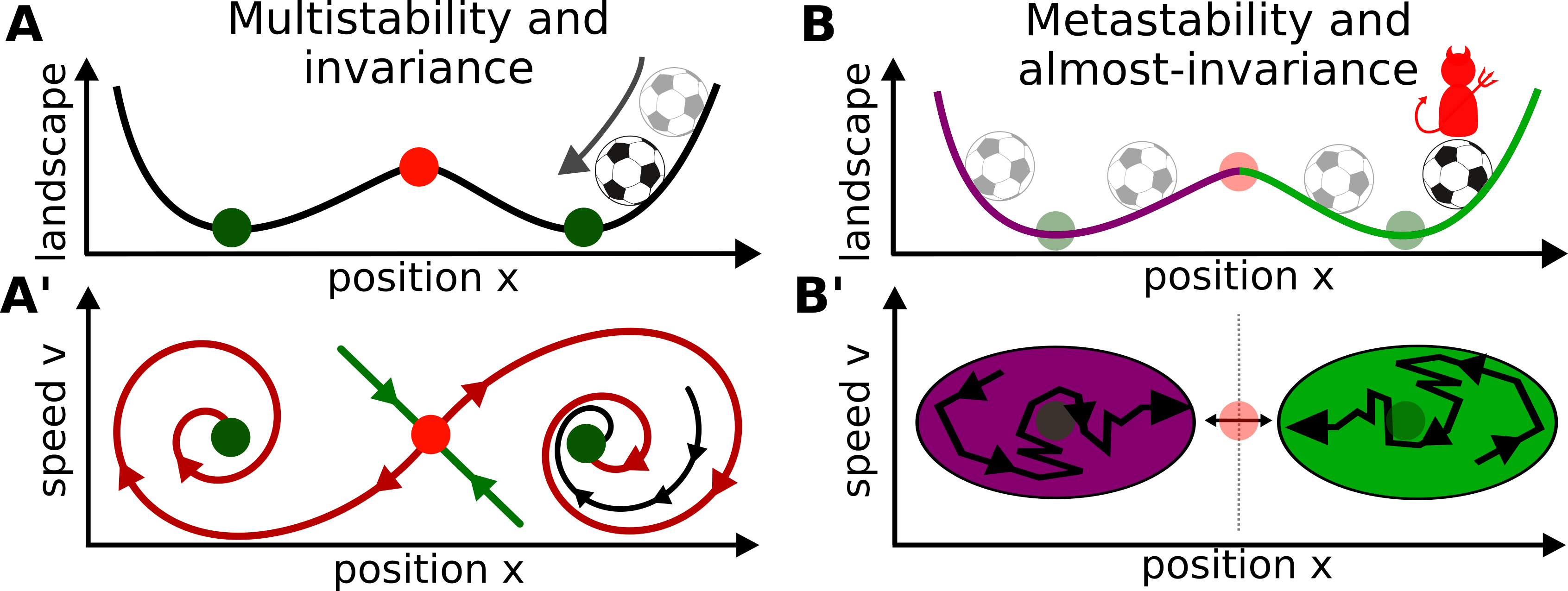}
    \caption{\textbf{Illustration of multistability versus metastability}. Panels A and A\supprime illustrate the behavior of a ball on a double-well landscape, representing a multistable system. The stable and unstable equilibria are represented by green and red circles, respectively. In A, the ball is shown on the landscape to converge to the equilibrium on the right, where it stays indefinitely (equilibria are invariant). In A\supprime, a possible corresponding trajectory is shown in state space. The green curves leaving the unstable equilibrium denote its stable manifold, while the red curves denote the unstable manifold. In B, we suppose there is a demon kicking the ball and thus introducing white noise to its dynamics. This now makes the ball switch eternally between the two wells, and effectively transforms the previous equilibria (not invariant anymore, represented as faded circles) into metastable regions (which are almost-invariant), represented illustratively in B\supprime by the purple and green ellipses. }
    \label{fig:concepts-dynamics}
\end{figure*}

For example, consider a ball moving in a landscape with two wells separated by a hill (cf. Fig. \ref{fig:concepts-dynamics}A). When released, the ball rolls downhill in this landscape. Appropriate variables defining the state of the ball at any given time are its position $x$ and velocity $v$, such that we can write the state $\textbf{s}$ of the ball as a two-dimensional vector $\mathbf{s} = (x,v)$. So the state space for the ball will have two dimensions: one for $x$, another for $v$, as represented in Fig.~\ref{fig:concepts-dynamics}A\supprime. We can initialize this system by releasing the ball from a certain height with some speed, defining an initial state $\mathbf{s}_0 = (x_0, v_0)$. 

This initial state, also called initial condition, is one point in state space. Once released, that ball will evolve in time, going through a continuous sequence of different states, revolving around the well until it eventually comes to rest at the bottom of one well. This sequence of states representing the movement of the ball is called a \textit{trajectory} of the system, and corresponds to a path in state space. Equivalently, a trajectory is a set of points (a set of states). There are infinitely many trajectories in state space; one of them is represented in Fig. \ref{fig:concepts-dynamics}A\supprime as the black curve.

Dynamical systems theory shows that the time evolution of the infinitely many trajectories is governed by the properties of certain structures present in state space. It then becomes crucial to understand these structures.

To start, notice that in the example the ball eventually converges to the bottom of one of the wells because it loses energy due to friction. In general, any system characterized by energy losses will converge to \textit{attractors}, which are attracting sets of points in state space. The state of the ball at rest on the bottom of each well is an equilibrium attractor, as the trajectories converge to that state and stay on it indefinitely. The equilibrium thus does not change under the time evolution of the system, and is called \textit{invariant}. If one were to periodically kick the ball, it may start to draw a periodic motion: instead of converging to a single point, it revolves around the well and repeats its motion every certain period. In state space this corresponds to a closed loop, also an invariant set. Other invariant sets are also possible, such as chaotic sets, in which the long-term behavior of a trajectory is highly sensitive to changes in its initial conditions.

Attractors are said to be locally stable in state space: any small perturbation away from an attractor leads to a trajectory that eventually returns to it. An important remark is that this double well system has two stable equilibria, i.e. two attractors, so it is called bistable (in general, for more attractors, it would be called \textit{multistable} \cite{feudel2008complex, feudel2018multistability}). To which of the two attractors the ball converges depends on its initial state. 

Another class of important structures are unstable sets. We have one such structure in the double well: the top of the hill. If a trajectory starts exactly on top of the hill, it will stay there for an infinitely long time; again, the top of the hill is invariant. If we perturb the ball, even slightly, it will leave the top and converge to one of the wells. 

The stability of invariant sets can be understood from a geometrical point of view in state space. While the stable equilibrium attracts trajectories from all directions near it, the unstable equilibrium has specific attracting and repelling directions. In this case, the unstable equilibrium is said to be of saddle type \cite{ott2002chaos}, in reference to saddles (e.g. a horse-riding saddle), which also have such attracting and repelling directions. These directions are called the stable and unstable manifolds of the equilibrium, respectively. Figure \ref{fig:concepts-dynamics}A\supprime shows the saddle equilibrium in a red circle, with its stable and unstable manifolds as green and red curves, respectively. Saddle-points, or in general saddle invariant sets, play an important role in several deterministic mechanisms leading to metastability, as we show in Sec.~\ref{sec:mechanisms}.

Before proceeding, we remark that an alternative introduction to many of these concepts is provided in Ref.~\cite{khona2022attractor}.

% state: $x \in M$; region of state space: $R = \{x\} \subset M$; regime: dynamical instantiation of $R$, as trajectories pass through it. invariance, stability

\subsection{Metastability and almost-invariance - definition}\label{sec:framework}
In the previous section we introduced a common notion of stability, which requires that a set is invariant, such that trajectories reaching it remain on it for infinitely long times. Now let us imagine a different scenario in the double well landscape: suppose the ball has a small demon that is constantly kicking it in randomly chosen directions with randomly chosen strengths, as illustrated in Fig.~\ref{fig:concepts-dynamics}B. This effectively adds noise to the ball´s evolution, creating a \textit{noisy} system. The landscape of course has not changed, so the \textit{noiseless} system is still the same. Therefore, the ball still tends to roll downhill, attracted towards the bottom. But the noise often kicks it away, repelling the ball from the bottom. As a result of this interplay between attracting and repelling tendencies, the ball is always in motion, revolving randomly around the bottom of one well, but never gets stuck to it. 

Eventually, by chance, the demon will give the ball successive kicks up the hill in such a way that it overcomes the top and falls onto the other well. How many kicks are needed depends, on average, on the noise strength: lower noise requires on average more kicks, such that the ball can stay around one well for very long epochs (see Ref.~ \cite{hangi1990reaction} for more details). 

So the ball with noise never converges to an attractor - instead, it moves for potentially very long times inside a region close to where the noiseless attractor is located, and eventually leaves. The noise transforms the ball´s dynamics from stable equilibria into metastable regimes, \textit{from multistability into metastability}, in the sense of long-lived but transient regimes that we discussed in Sec.~\ref{sec:common-thread} \cite{hangi1990reaction,pisarchik2014control, arecchi1985generalized, kraut1999preference, kraut2002multistability, kraut2003enhancement, feudel2008complex}. This idea is illustrated in Figs.~\ref{fig:concepts-dynamics}B and B\supprime.  It should be noted that these remarks are valid as long as the noise strengths are not large enough to create purely noisy trajectories, which just diffuse through the state space.

In the noiseless system, the bottom of each well forms an invariant set. In the noisy system, we have seen that two regions emerge in which the ball spends a long time. These regions are \textit{almost-invariant}, in the sense that trajectories inside them have a high probability of remaining inside as they evolve \cite{froyland2005statistically, dellnitz2003congestion}. An almost-invariant region $R$ can be defined such that the probability of staying inside it after some time evolution is large enough that trajectories stay inside for long periods of time. The higher the probability, the longer they stay.  These concepts have been nicely formalized in previous works \cite{froyland2005statistically, dellnitz2003congestion}, so we refer the reader to those references for more details. We provide a brief summary in the Supplementary Material.

Importantly, note that noise is, in general, not necessary for generating almost-invariant regions - as we see later in Sec.~\ref{sec:mechanisms}, there are also several mechanisms that lead to almost-invariant regions in deterministic, autonomous systems. The case with noise was used here only as a simple concrete example.

Therefore, we can say in general that a region $R$ of state space is metastable if it is almost-invariant, that is, if trajectories, once in $R$, tend to stay in $R$ for a long time before leaving. This is a key point in our Perspective: \Emph{metastability is observed through \textit{long-lived but transient} (metastable) \textit{regimes}, and underlying any metastable regime is an \textit{almost-invariant} (metastable) \textit{region} of state space}.  

This view formalizes the notion of long-lived but transient regimes and reformulates them into properties in state space, associating a dynamical regime to its corresponding region, and associating that region to an interpretable quantity, namely the probability of trajectories remaining inside. The concept of almost-invariance is crucial for this, and serves as a theoretical tool to quantitatively characterize metastability.

In some applications, one might be concerned with the minimum duration of metastable regimes, i.e., how long regimes need to be in order to be metastable. This question requires an arbitrary decision on a threshold, which cannot be objectively decided in general; it depends on the specific application. The advantage of this definition is that it is agnostic to a minimum duration and thus flexible to this decision. Another advantage is that the decision can be made not in terms of a duration, but in terms of a probability, which may have a clearer interpretation. 

Beyond their characterization, the identification of metastable regimes is also an important problem, which has seen several solutions in the neuroscience literature, as discussed in Sec.~\ref{sec:observations}. An interesting approach is through optimization procedures that aim to maximize the probability of trajectories remaining inside the almost-invariant regions \cite{froyland2003detecting, froyland2005statistically, dellnitz2003congestion}. This can identify maximally almost-invariant regions, on which regimes will be maximally long, on average. These regions have been shown for some systems to have a particular geometrical property: they are partially separated in state space by invariant manifolds, which gives further insight into their structure and localization \cite{froyland2009almost}. 
The technical details of how this identification can be done for observed time-series, however, go beyond the scope of this Perspective, and is the subject of ongoing research \cite{froyland2018robust}. 

Further, it is worth remarking that metastability of a regime does not fully define its dynamics. As we see in the next section, metastable regimes can be, for instance, periodic or chaotic, generated by deterministic or by noisy mechanisms, etc. A deeper analysis is required to fully characterize metastable regimes. 

%An interesting question to be investigated is how the methods previously applied in neuroscience relate to these optimization methods.

\subsection{Dynamics-based mechanisms for metastability} \label{sec:mechanisms}

The definition of metastability encompasses the observations and formulations previously reported in neuroscience. Furthermore, and crucially, it also allows one to study concrete mechanisms that generate metastable dynamics. These are known from dynamical systems theory. Some have been proposed in the context of metastability \cite{graben2019metastable, cavanna2018dynamic, brinkman2022metastable}, but others have not yet so far, to our knowledge.

Before we proceed to the mechanisms, we remark that around any metastable region there is a \textit{coexistence of attracting and repelling tendencies} \cite{tognoli2014metastable, kaneko2003chaotic}. The ability of metastable regions to retain trajectories is due to their attracting tendencies, and their finite duration is due to their repelling tendencies that push the trajectories away - for instance, in the double-well example, think of the attraction towards either of the two wells versus the repulsion due to the noise.

\begin{figure*}[hbt]
    \centering
    \includegraphics[width=\textwidth]{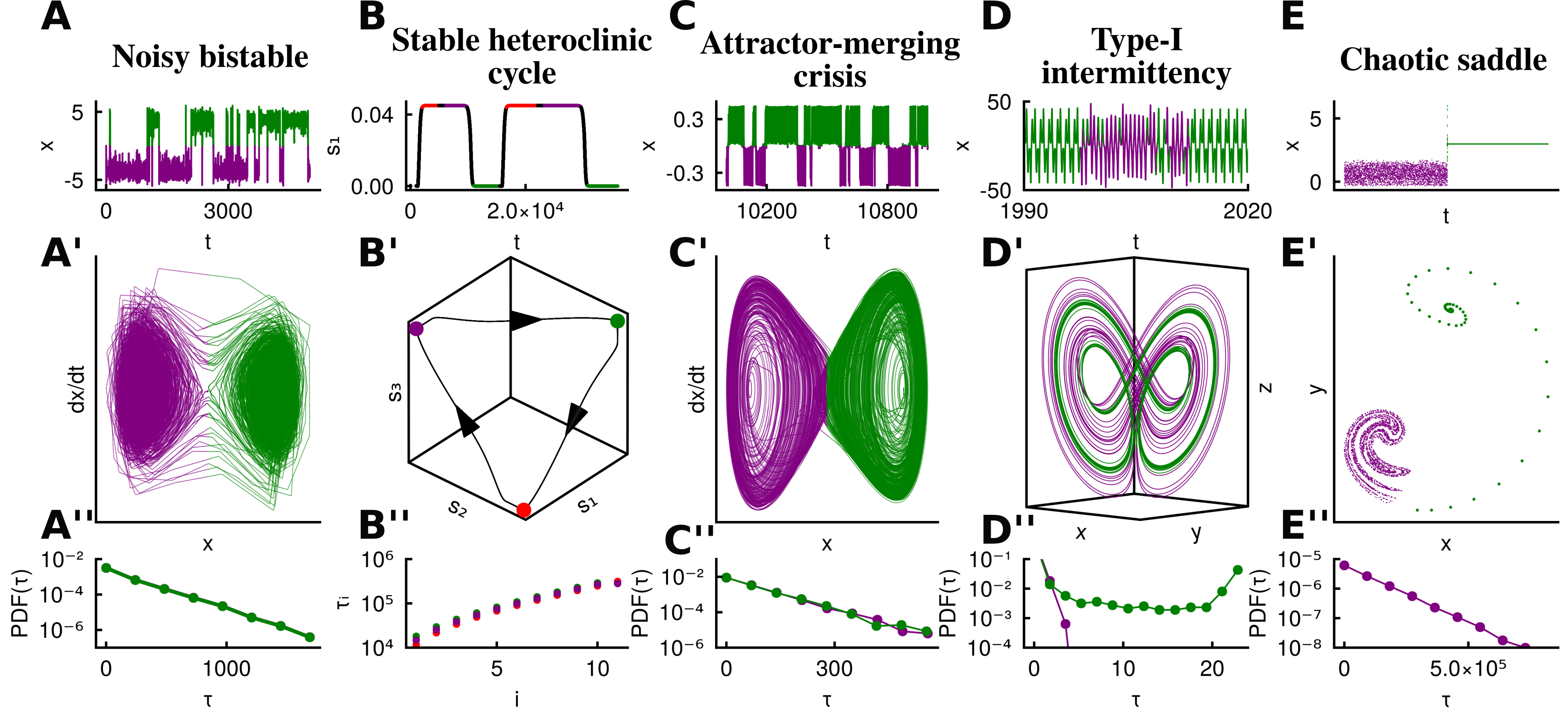}
    \caption{\textbf{Dynamics-based mechanisms of metastability.} Each column corresponds to a distinct mechanism. The first row (A-E) shows a representative time-series, with alternations between metastable regimes, each receiving a distinct color. The second row (A\supprime-E\supprime) shows the state space of each system with the trajectory corresponding to the time-series above. The third row (A\ssupprime-E\ssupprime) shows the distribution of residence times in each metastable region for several trajectories. Further details are discussed in the main text and Supplemental Material.}
    \label{fig:mechanismsmetastability}
\end{figure*}

The double-well landscape with noise is a simple example of the general behavior of \Emph{noise in a multistable system}. As we have discussed before, trajectories spend considerable time around one well, then hop to the other well, and repeat this behavior indefinitely. This phenomenon is called \textit{attractor hopping} \cite{kraut2002multistability}: the system now has two metastable regimes, each corresponding to the dynamics around one well. In this sense, \textit{noise replaces the stable equilibria with metastable regions}. The regimes can be seen in the time-series of the ball's position $x$ in Fig. \ref{fig:mechanismsmetastability}A and the regions in the ball's state space, spanned by its position $x$ and velocity $v$, shown in Fig. \ref{fig:mechanismsmetastability}A\supprime. The times spent on each metastable region (called residence times or dwell times) follow a probability distribution, such as an exponential distribution for Gaussian white noise \cite{hanggi1986escape} (Fig. \ref{fig:mechanismsmetastability}A\textsuperscript{\everymodeprime\everymodeprime}). This mechanism of noise-induced transitions in multistable systems has been proposed for several observations of metastability in the brain \cite{brinkman2022metastable, hudson2017metastability}.

We introduced metastability with the example of a noisy system in Figs.~\ref{fig:concepts-dynamics}B-B\supprime and \ref{fig:mechanismsmetastability}A-A\ssupprime, but metastability can also occur without any noise or external input, due to several possible \textit{deterministic} mechanisms. We concentrate now on some of these, starting with a \Emph{stable heteroclinic cycle}, which is demonstrated here by a firing-rate model derived from three synaptically-coupled Hodgkin-Huxley neurons \cite{ashwin2011criteria}. Figure \ref{fig:mechanismsmetastability}B shows the time-series of a trajectory on such a cycle, performing a sequential alternation between three distinct regimes, each with a unique color. These regimes correspond to the vicinity of specific points in state space, as highlighted in Fig.~\ref{fig:mechanismsmetastability}B\supprime. They are saddle-points, like the red circle shown in Figs.~\ref{fig:concepts-dynamics}A-A\supprime, which have coexisting attracting directions (stable manifold) and repelling directions (unstable manifold). The stable manifold of one saddle is connected to the unstable manifold of the next, in what is called a heteroclinic connection \cite{ott2002chaos}. Furthermore, this cycle is stable: trajectories approach one of the saddle-points along its stable manifold and potentially spend a long time near it - then, they depart along its unstable manifold to the next saddle, and so keep cycling through them, converging ever closer to the saddles and their manifolds. This can be seen in the increasing residence times in Fig.~\ref{fig:mechanismsmetastability}B\ssupprime. Trajectories in the neighborhood of each saddle-point stay inside for considerable durations, so the neighborhood is a metastable region (even though the saddle-point itself is invariant and unstable).

Importantly, it can be shown that such heteroclinic cycles can globally attract trajectories \cite{nowotny2007dynamical} and are also conserved under small parameter changes (they are structurally stable) \cite{rabinovich2008transientcognitive}.  Moreover, it has been shown in neural networks that heteroclinic cycles depend on the inputs to the neurons, such that the cycle is sensitive to the stimulus to the network \cite{rabinovich2001dynamical}. Once the stimulus is applied, and the cycle is defined, it is robust against noise. So this mechanism can allow for robustness to noise while keeping sensitivity to inputs \cite{rabinovich2001dynamical, rabinovich2008transientcognitive}. This, on top of the cyclic behavior, means that heteroclinic cycles are important structures to perform sequential \textit{computations} \cite{fonollosa2015learning}.
% An example of this is for chunking dynamics, whereby information to be stored in memory is represented a phone number, e.g. $858-535-22-30$, is separated into $4$ chunks: $848$, $535$, $22$, and $30$. Trajectories representing this number cycle through the chunks as $848 \to 535 \to 22 \to 30$. Each chunk by itself is composed of sequences of saddle points representing the digits, e.g. $8 \to 5 \to 8$. In this way, sequential memories can be encoded by heteroclinic cycles of heteroclinic cycles of saddle points \cite{fonollosa2015learning}}. 
An example of this is chunking dynamics, in which a trajectory coding for a memory alternates along a heteroclinic cycle composed of further heteroclinic cycles, each corresponding to chunks of a memory \cite{fonollosa2015learning}. As such, stable heteroclinic cycles have also been proposed as the mechanism behind experimental observations \cite{rabinovich2008transientcognitive, rabinovich2008transientdynamics, rabinovich2012information} and theoretically hypothesized as a mechanism for cognition \cite{fonollosa2015learning, rabinovich2014chunking}. Heteroclinic cycles can be extended to heteroclinic networks; for a review, see \cite{hildegard2023heteroclinic}. 

% Now we move to another example, which involves a \textit{chaotic attractor} that originates from an \textbf{attractor-merging crisis}. \kl{In this case, the chaotic attractor can be decomposed into two distinct sub-regions, each of which acts almost like a separate chaotic attractor: trajectories spend on average a long time on one sub-region before switching to another sub-region, repeating this process indefinitely. Figures \ref{fig:mechanismsmetastability}C-C\supprime illustrate this intermittency in a time-series and state space, respectively, with the metastable regimes or regions colored green and purple. The transitions between the regions are mediated by a particular periodic trajectory of type saddle (having stable and unstable manifolds), which therefore acts as the gateway between the regions - trajectories on one sub-region approach it and are taken to the other sub-region. It thus acts as the repelling tendency mediating the transitions between the sub-regions. Geometrically, the attractor-merging crisis occurs through a parameter change that causes two separate chaotic attractors to collide with a saddle periodic orbit, and become connected in a single, larger, chaotic attractor}.
Now we move to another example, which involves a \textit{chaotic attractor} that originates from an \textbf{attractor-merging crisis} \cite{grebogi1987critical}. In this case, the chaotic attractor can be decomposed into two distinct sub-regions. For a different parameter value of the system, these sub-regions are separate chaotic attractors. Starting from some critical parameter value, they merge together into a combined chaotic attractor, with a specific pathway connecting them. Then, trajectories in each sub-region oscillate chaotically for a long time, on average, before they switch to the other sub-region, and repeat this process indefinitely. Figures \ref{fig:mechanismsmetastability}C-C\supprime illustrate these two intermittent regimes in a time-series and their corresponding regions in state space, respectively, colored green and purple in both. The specific pathway guiding the transitions between each sub-region is a particular periodic trajectory of type saddle (having stable and unstable manifolds) - trajectories on one sub-region approach this trajectory and then are taken to the other sub-region. It thus acts as the repelling tendency mediating the transitions between the sub-regions.
The probability distribution of residence times on each of these metastable regions is exponential (Fig. \ref{fig:mechanismsmetastability}C\ssupprime). 

It is interesting to note that the trajectories for this mechanism look similar to those for the noisy bistable systems (compare Figs. \ref{fig:mechanismsmetastability}A and \ref{fig:mechanismsmetastability}C). The key difference is that the transitions in the former are caused by noise, and in the latter they are caused purely by the deterministic dynamics of the system on the attractor, with the mediating saddle periodic trajectory. Still, this highlights the often apparent similarity between noisy and chaotic dynamics, and the need for deeper analysis to distinguish between both \cite{boaretto2021discriminating}. 

Similar dynamics occur as a result of another mechanism, an \textbf{interior crisis} \cite{grebogi1983crises, ott2002chaos}. Also in this case, trajectories intermittently switch between two sub-regions of the same chaotic attractor. But the underlying structures are different from the attractor-merging crisis: a change of parameters causes a chaotic attractor to connect to a chaotic set that is unstable (a chaotic saddle, which we see more in Fig. \ref{fig:mechanismsmetastability}E-E\ssupprime). This connected chaotic attractor now includes both regions, but trajectories stay inside each sub-region for long times, before switching to the other. These transitions are again mediated by a special pathway, a periodic trajectory of saddle type, which acts as the repelling tendency. The distribution of times on the region occupied by the original chaotic attractor is again exponential. See Ref.~\cite{ott2002chaos} for more details.

Another example of a chaotic attractor that can be decomposed into two metastable regions occurs in \Emph{type-I intermittency}. We show in Fig.~\ref{fig:mechanismsmetastability}D an example in a 3-dimensional system. The trajectory alternates between metastable regimes: one with clear chaotic dynamics (in purple) and another with seemingly periodic dynamics (green). Looking at the state space in Fig.~\ref{fig:mechanismsmetastability}D\supprime, we note that the purple regime corresponds to a region that looks clearly chaotic (fractal-like structure), while the green regime corresponds to a region that looks like a periodic orbit (a closed loop). Trajectories can spend a considerable amount of time on the green region, looking periodic. When they leave it, they can then spend a long time in the purple region, leaving it when they approach the green region again.
Mechanistically, the green region is actually the \textit{ghost} of a stable periodic orbit that exists for different parameter values \cite{pomeau1979intermittency, pomeau1980intermittent, strogatz2002nonlinear}. That stable periodic orbit is destroyed by a collision with an unstable periodic orbit at some critical parameter value, leaving behind only the ghost, which is not an invariant structure \cite{medeiros2016trapping}. The bifurcation giving rise to this behavior is a \textit{saddle-node bifurcation of periodic orbits}. It can also occur analogously for equilibria, and we refer the reader to Refs.~\cite{koch2023beyond, sussillo2013opening} for further work and illustrations. Importantly, the two metastable regions in our example belong to the same chaotic attractor, and the trajectories near the ghost merely \textit{appear} to be periodic. 
Further, the time spent near the ghost can be considerably long, but is finite for any fixed parameter \cite{pomeau1980intermittent} (see the distribution of times in Fig. \ref{fig:mechanismsmetastability}D\ssupprime).

The type-I intermittency mechanism, along with the two crises discussed before, are three of many concrete instances of an idea proposed by Friston \cite{friston2000transients}, who argues that different metastable regimes can correspond to different sub-regions of a same invariant set (usually attracting). 

An important remark is that this mechanism does not require a chaotic attractor, as it occurs whenever a saddle-node bifurcation is nearby in parameter space. The chaotic attractor in the example only acts to re-inject the trajectories that leave the neighborhood of the ghost back onto it, creating a repeatable metastability. This re-injection can also occur due to noise, not only chaos \cite{medeiros2016trapping}. But if there is no re-injection mechanism, nearby trajectories will pass the neighborhood of the ghost only once, and then this is an example of non-repeatable metastability.

This saddle-node mechanism has been proposed by Kelso and colleagues as a mechanism for metastability in the brain \cite{kelso1991an, tognoli2014metastable}. Ghost structures have also been shown to play a role in the dynamics of recurrent neural networks performing particular computations \cite{sussillo2013opening}. Further, similarly to the idea of cycles of saddle-points (Figs. \ref{fig:mechanismsmetastability}B-B\ssupprime), cycles of ghost structures can also be constructed, and have been shown to generate sequences of metastable regimes that are robust in the presence of noise, having potential advantages over cycles of saddle-points \cite{koch2023beyond}. 

% It is another example of the general proposal by Friston \cite{friston2000transients}, since the trajectories here live on a chaotic attractor, which can nevertheless be separated in two distinct dynamical regimes: nearly-periodic (green) and chaotic dynamics (purple). 

Another mechanism that leads to metastable regimes is due to a \Emph{chaotic saddle}. A chaotic saddle is similar to a chaotic attractor, with the distinction that it also has repelling directions - it is thus the analogue of a saddle equilibrium for a chaotic set \cite{lai2009transient}. As a consequence, trajectories can stay near the chaotic saddle for potentially very long times, displaying chaotic dynamics, but are eventually expelled from it \cite{yorke1979metastable}. Figure \ref{fig:mechanismsmetastability}E shows a trajectory switching from a regime near a chaotic saddle (in purple) to one on an equilibrium (in green); the corresponding regions in state space are shown in Fig. \ref{fig:mechanismsmetastability}E\supprime. Note that the trajectories leave the chaotic saddle because they are not on it exactly; they are only near it. These are the typical trajectories, which are the ones numerically and experimentally observed. They form a metastable region even though the chaotic saddle itself is invariant and unstable. Ultimately, the trajectories converge to the stable equilibrium and remain on it infinitely long.
The residence times for different initial conditions near the chaotic saddle follow an exponential distribution \cite{grebogi1983crises} (Fig.~\ref{fig:mechanismsmetastability}E\ssupprime).
% We remark that trajectories on the chaotic saddle resemble the ignition behavior leading to conscious access shown in Ref. \cite{dehaene2005ongoing}.

A chaotic saddle has been reported in networks of spiking neurons \cite{ansmann2016selfinduced}, in which, interestingly, it can be subdivided into three distinct dynamical regimes, corresponding to three distinct sub-regions, which are all intermittently visited for considerable durations. Therefore, also every sub-region of that chaotic saddle can be considered metastable. That is thus an example of \textit{hierarchical metastability}, in which metastable regions contain inside more metastable regions, and the dynamics progresses through a hierarchy of metastable regimes \cite{cavanna2018dynamic}. Chaotic saddles have also been observed in other networks of spiking neurons \cite{keplinger2014transient, lafranceschina2015impact, hartle2017transient, kaminker2019alternating}, and may play a key role in synchronization processes \cite{medeiros2018boundaries, medeiros2019state}.

There are several other dynamics-based mechanisms for metastability beyond the ones discussed previously. One example is \Emph{on-off intermittency}, which is common, for instance in systems that can completely synchronize \cite{ashwin1994bubbling, ott1994blowout}, and occurs during transitions from synchronization to desynchronization induced by changing the coupling strength between units. During these transitions, systems with on-off intermittency have trajectories that intermittently switch from a synchronized regime to a desynchronized one and back. The mechanism for this intermittency, and its statistical scaling, has been studied in detail \cite{ platt1993onoff, heagy1994characterization, hammer1994experimental, cenis1997symmetry}. It is one out of many possible mechanisms that can realize the coexistence of integration (synchronization) and segregation (desynchronization), often mentioned as a crucial need for the brain \cite{fingelkurts2001operational, tognoli2014metastable, tognoli2014enlarging, deco2015rethinking}. It is also a possible mechanism for the intermittent switching between UP and DOWN states exemplified in Fig. \ref{fig:observationsmetastability}C. Furthermore, evidence for on-off intermittency has been reported for EEG data of non-convulsive paroxysmal (high-amplitude) activity in rats with genetic absence epilepsy \cite{hramov2006onoff}. The study identified a power-law distribution of residence times characteristic of on-off intermittency in regular brain activity, which alternated with the epileptic paroxysms.
A very similar mechanism is in-out intermittency \cite{ashwin1999transverse, ashwin2001influence,saha2018characteristics, saha2017extreme}. In addition, evidence for yet another type of intermittency, known as type-III intermittency \cite{pomeau1979intermittency}, has been reported in \cite{velazquez1999type}.

Metastability can also manifest as \Emph{mixed-mode oscillations}, in which trajectories alternate between regimes of small and large amplitude oscillations \cite{desroches2012mixed}. There are various mechanisms for mixed-mode oscillations, which can involve forcing the system to induce changes in parameters or through an adaptation variable that enables the switching between the two modes \cite{desroches2012mixed, rotstein2014mixedmode}. 
A particular example of mixed-mode oscillations is \emph{bursting} behavior occurring in neurons, characterized by quick firing of spikes followed by a quiescent period (as in Fig. \ref{fig:observationsmetastability}F). Bursting is a ubiquitous mode of firing in neurons \cite{fox2015bursting}, and can have important functional roles due to its ability to generate more robust responses \cite{swadlow2001impact}. A simple model displaying bursting is the Hindmarsh-Rose neuron \cite{hindmarshmodel1984}. In it, an adaptation current causes the switching between the tonic spiking and the silence \cite{hindmarshmodel1984}. The whole trajectory is on a stable limit cycle, but can still be decomposed into the tonic spiking and the silence, each of which can be considered metastable. This is yet another example of the idea proposed by Friston in \cite{friston2000transients}. Beyond this model, bursting behavior occurs in systems with a period-adding cascade \cite{kartanak2014route}.  
Mixed-mode oscillations can also occur as neuronal avalanches, which are cascades of activity occurring in varying sizes and durations, and across different scales, seen in local-field potential activity \cite{beggs2003neuronal}. An example of a dynamics-based mechanism generating avalanches has been recently elucidated in \cite{contreras2023scale}. The onset of these avalanches may be associated with critical phenomena in the brain (see Ref. \cite{girardi2021brain} for a recent review).

% Furthermore, in neuronal networks, spontaneous local field potential activity cascades across various scales within the system, giving rise to neuronal avalanches of varying sizes and durations \cite{beggs2003neuronal}, providing another example of metastability. An example of a dynamical mechanism generating avalanches has been recently elucidated in \cite{contreras2023scale}. As a branching process, the onset of these avalanches may be associated with critical phenomena in the brain (see Ref. \cite{girardi2021brain} for a recent review).}

% A similar important behavior is \Emph{mixed-mode oscillations}, in which trajectories alternate between periods of small amplitude (e.g., sub-threshold) and large amplitude oscillations (e.g., spikes) \cite{rotstein2014mixedmode}. There are various mechanisms for mixed-mode oscillations, which can involve either forcing the system to induce changes of parameters or through an adaptation variable that enables the switching between the two modes \cite{desroches2012mixed, rotstein2014mixedmode}.

Moreover, an important example of metastable behavior is \Emph{chaotic itinerancy} \cite{kaneko2003chaotic}. This phenomenon has also been discussed in the context of neuroscience \cite{freeman2003evidence, tsuda2009hypotheses, tsuda2015chaotic}, and can include several specific mechanisms, and thus is also an umbrella term, but more specific than metastability as we define it here.

\subsection{Subtypes of metastability} \label{sec:subtypes}
The definition of metastability proposed in Sec.~\ref{sec:framework} acts as an umbrella term encompassing several distinct observations, formulations, and mechanisms. More specific formulations, which restrict the phenomenon to particular contexts, can be defined as subtypes of metastability. 

As a first example, we may think of \Emph{spontaneous and driven} metastability (illustrated in Figs.~\ref{fig:definitionsmetastability}G-G\supprime respectively). They differ in regards to how transitions between metastable states are generated. In \textit{spontaneous} metastability, the mechanism behind transition is intrinsic to the system, and thus occurs without the need of external perturbations - such as noise, or stimuli. This is the case for Figs. \ref{fig:mechanismsmetastability}B-E\ssupprime. Driven metastability is the opposite: transitions occur because of external perturbations, as in Fig. \ref{fig:mechanismsmetastability}A-A\ssupprime. Each subtype has distinct but important functional roles: spontaneous metastability enables transitions between regimes without the expenditure of additional energy to stimulate the transitions \cite{tognoli2014metastable}, and guarantees that a system does not get stuck in the same regime \cite{ito2008dynamics}; driven metastability is important for direct control of the transitions, such as in stimulus-evoked scenarios.

A second important pair of subtypes is \Emph{repeatable versus non-repeatable metastability} (Figs. \ref{fig:definitionsmetastability}H-H\supprime). These are characterized by whether the metastable regimes eventually repeat in the time series or not. If the metastable regimes repeat along observations, the metastability is repeatable; if they do not, it is non-repeatable. The case of non-repeatable regimes is sometimes called metastability en route to ground state \cite{brinkman2022metastable}. Note that the observations in Fig. \ref{fig:observationsmetastability} and the mechanisms in Fig. \ref{fig:mechanismsmetastability} are already classified into these subtypes.

\section{Conclusions and outlook}
In this Perspective we have provided a discussion on metastable brain dynamics in a way that is consistent across observations and theory in both neuroscience and dynamical systems. With this, we have looked into several dynamics-based mechanisms that generate metastability, and compared them to obtain insights into their commonalities, distinctions, and connections to previous literature.

The first step was to extract a general definition of metastability from the literature: metastable regimes are long-lived but transient. This serves as an umbrella term encompassing the other formulations in the literature, is useful to interpret observations, and readily connects to dynamical systems theory. We then looked deeper into how these metastable regimes behave in state space. To achieve this, we introduced the notion of almost-invariant regions of state space, characterized by a high probability (smaller than one) that trajectories, once inside them, remain inside \cite{froyland2005statistically, dellnitz2003congestion}. Trajectories can enter an almost-invariant region, stay inside for a long time, and eventually leave. While inside such a region, trajectories exhibit time-series containing long-lived but transient epochs with unique dynamical properties, i.e. metastable regimes. The dynamics of the regimes corresponds to the dynamics in the regions (e.g., periodic or chaotic). Thus, almost-invariant regions, which we identify as metastable regions, underlie the dynamics of metastable regimes. 

With this view, we have then discussed the \textit{dynamics-based mechanisms} for metastability, i.e. how metastable regions and regimes are generated, and how they behave. Dynamical systems theory provides several concrete possibilities. Some of these have been previously discussed in the neuroscientific context \cite{graben2019metastable, cavanna2018dynamic, brinkman2022metastable}, but some are also new, to the best of our knowledge. Though not a comprehensive list, we believe it serves as a good starting point for researchers both in neuroscience and dynamical systems intending to study metastability. 

Looking at so many different mechanisms together provides several insights. To start, it provides concrete examples of metastable regimes corresponding to trajectories visiting metastable (almost-invariant) regions. Further, it becomes clear that around each metastable region there is a \textit{coexistence of attracting and repelling tendencies in state space}, as has been discussed in Refs. \cite{tognoli2014metastable, kaneko2003chaotic}. These realizations further validate the definition, which is general but still meaningful to allow for common underlying principles.

With this rich list of possible mechanisms, we have been able to provide several concrete instances of an idea that was abstractly proposed by Friston in Ref.~\cite{friston2000transients}: that the same attractor can have multiple sub-regions, each corresponding to one metastable regime. An interesting instance of this has been reported in Ref.~\cite{ansmann2016selfinduced}, in which networks of spiking neurons display a chaotic saddle (its neighborhood a metastable region) that can be further subdivided into three sub-regions, each metastable and corresponding to different patterns of activity. 

We have also suggested more dynamics-based mechanisms that can account for the important behavior of coexistence of segregation and integration: besides the important saddle-node bifurcation of fixed points that often occurs \cite{tognoli2014metastable}, we also mention in-out and on-off intermittency as mechanisms that lead to variability of synchronization \cite{ashwin1994bubbling, saha2018characteristics}. 

The solid theoretical background offered by the idea of almost-invariant regions allows us to provide objective answers to issues debated in the literature. The distinction between attractors and metastable regions, and hence between \textit{multistability and metastability} becomes clear. As discussed in Sections~\ref{sec:stability_and_invariance} and \ref{sec:framework}, attractors are invariant: trajectories on them remain on them infinitely long; meanwhile, metastable regions are only almost-invariant: trajectories eventually leave them. 
A system is called multistable when it has multiple coexisting attractors for fixed parameters, a typical behavior in nonlinear dynamical systems \cite{feudel2008complex, feudel2018multistability}. In a multistable system, trajectories converge to one of the attractors, selected by their initial conditions, and remain on this attractor forever. There are no transitions between attractors in a multistable system. If external perturbations such as noise are introduced to a multistable system, the trajectories do not stick to attractors anymore, but can hop between them \cite{hangi1990reaction}. In this case, multistability is \textit{substituted} by metastability \cite{pisarchik2014control}. An important remark is that, when the noise is too strong, it can overpower any dynamics of the system and then there is neither multistability nor metastability observable, only a random motion \cite{kraut2002multistability}.
Crucially, however, stability (and multistability) can \textit{co-occur} with metastability in autonomous systems, without noise. For instance, an attractor can be in some cases decomposed into two or more metastable regions, as discussed previously and in Sec.~\ref{sec:mechanisms}. Then, trajectories remain eternally inside an attractor, but with transitions between its metastable regions.  
% If a system has multiple attractors for fixed parameters, a typical behavior in nonlinear dynamical systems, it is multistable \cite{feudel2008complex, feudel2018multistability}. In a multistable system, trajectories converge to one of the attractors, selected by their initial conditions, and remain on the attractor forever. There are no transitions between attractors in a multistable system. If external perturbations such as noise are introduced to a multistable system, the attractors cease to be invariant regions. If the noise is not too strong, the attractors are substituted by metastable regions, and trajectories can hop between these regions. If the noise is too strong, no patterns are seen \cite{kraut2002multistability}, and there is neither multistability nor metastability observable, only a random motion. 
% In the case of sufficiently weak noise in a multistable system, the multistability is \textit{substituted} by metastability. However, in other circumstances, without noise, stability (and multistability) can \textit{co-occur} with metastability. For instance, an attractor can be in some cases decomposed into two or more metastable regions, as discussed previously and in Sec.~\ref{sec:mechanisms}. Then, trajectories remain eternally inside an attractor, but with transitions between its metastable regions.  

Another point that is often debated is whether both \textit{spontaneous and noise-induced} cases should be considered as metastable \cite{kelso2012multistability, cavanna2018dynamic, hudson2017metastability}. Here, we provide an objective answer: since there are almost-invariant regions underlying the dynamics in both cases, both are metastable. This is exemplified through the motion in the double-well landscape with noise (Figs.~\ref{fig:mechanismsmetastability}A-A\ssupprime) and the intermittency due to an attractor-merging crisis (Figs.~\ref{fig:mechanismsmetastability}C-C\ssupprime), in which also the phenomenology in the time-series, in state space, and in the distribution of residence times is strikingly similar. This means that many dynamics-based mechanisms may underlie the same observation. Therefore, the characterization of regimes as metastable is only part of the story: they possess other dynamical properties that also need to be studied. Developing methods to achieve this and properly distinguish between various mechanisms is still a subject of future research.

Understanding dynamics-based mechanisms is an important first step for \textit{predicting} transitions between metastable regimes. This is particularly important in the context of seizure prediction, an area that has deservedly received a lot of attention but which still has many open questions \cite{kuhlmann2018seizure}. Knowledge of the dynamics-based mechanism underlying transitions from normal brain activity to seizures has been recognized as a key development in the field \cite{kuhlmann2018seizure}, but that alone may not be enough: part of the difficulty may be the co-existence of several mechanisms, not just one. A more comprehensive understanding of the mechanisms for metastability, extending the ideas proposed here, could prove useful to address these difficulties through a better understanding of the precursor dynamics of transitions \cite{rings2019traceability}. A possible future path can also aim to determine mechanisms underlying observations through the use of techniques to identify system equations from data \cite{voss2004nonlinear, brunton2016discovering, tabar2019analysis, anvari2016disentangling, jacobs2023hypersindy}. 

Finally, we believe it is worth mentioning that transient or metastable dynamics have been recognized as important mechanisms for \textit{computations} in the brain \cite{fonollosa2015learning, lacamera2019cortical, wang202150, laje2013robust, mazor2005transient, sussillo2013opening} and other biological systems \cite{nandan2023non, nandan2022cells}. Research has shown that systems can successfully perform computations with attractors \cite{khona2022attractor} or with transient regimes \cite{mazor2005transient,buonomano2009state, laje2013robust, khona2022attractor, benigno2023waves, budzinski2023an, liboni2023image}. This, allied with the observations of complex metastable dynamics in the brain \cite{brinkman2022metastable}, brings up the question: what advantages can spontaneous metastable regimes provide for computations, in contrast to stable regimes? This is a crucial question that needs to be addressed in future works, and may use the ideas proposed here as a starting point.

Therefore, we believe that the perspective we provide here connects observations, formulations, and mechanisms of metastability in neuroscience with the goal of understanding metastable brain dynamics.

\section*{Acknowledgments}
We would like to thank Klaus Lehnertz, James A. Yorke, Niccolò Zagli, Nicolás Rubido, and Péter Koltai for inspiring discussions. In particular we would like to deeply thank Klaus Lehnertz for comments on this manuscript. K.L.R. was supported by the German Academic Exchange Service (DAAD). E.S.M and U.F. acknowledge the support by the Deutsche Forschungsgemeinschaft (DFG) via the project number 454054251. B.R.R.B. acknowledges the support of the São Paulo Research Foundation (FAPESP), Brazil, Proc. 2018/03211-6 and 2021/09839-0. This work was partially supported by BrainsCAN at Western University through the Canada First Research Excellence Fund (CFREF), the NSF through a NeuroNex award (\#2015276), the Natural Sciences and Engineering Research Council of Canada (NSERC) grant R0370A01, and the Western Academy for Advanced Research. R.C.B gratefully acknowledges the Western Institute for Neuroscience Clinical Research Postdoctoral Fellowship.

\section*{Supplementary material}
\subsection{Models}
The code used to generate Figure 4 in the main text is available in the GitHub repository \cite{rossi2022repository}. The data used can be made available upon request. All the code is done in the Julia computational language \cite{bezanson2017julia}; integration was done with the package DifferentialEquations.jl \cite{rackauckas2016differential}, with the aid of packages DynamicalSystems.jl \cite{datseris2018dynamical} and DrWatson.jl \cite{datseris2020drwatson}. Plots were made with Makie.jl \cite{danisch2021makie}.

\subsubsection{Noisy bistable system}
The noisy bistable system, also known as the noisy Duffing oscillator \cite{strogatz2002nonlinear}, is given by:
\begin{align}\label{eq:duffing}
    dx &= v + \eta_1 dW\\ 
    dv &= -ax^3 + bx -c -dv + \eta_2 dW,
\end{align}
with $dW$ describing a white Gaussian noise. The equations describe the evolution of a particle on a double-well (quartic) potential $U(x) = ax^4/4 -bx^2/2 +cx$ being periodically driven and with noise.
The parameters used were $a=0.5$; $b=8.0$, $c=0.0$, $d = 0.2$, $\eta_1= \eta_2 = 0.18$, with initial condition $(0, 0)$. 

% Each well corresponds to a fixed point in this system. Without noise, trajectories would converge to one of the fixed points and stay there indefinitely, as they would be stable. But the noise kicks trajectories away from the fixed points, such that, the fixed point now become metastable. For noise amplitudes that are not too high, the trajectory keeps alternating between dwelling near one (now metastable) fixed point and dwelling near the other, as shown in Fig. \ref{fig:observationsmetastability}(a). 

\subsubsection{The heteroclinic cycle}
The heteroclinic cycle occurs in a rate model derived from Hodgkin-Huxley type neurons with synaptic coupling \cite{ashwin2011criteria}. The equations are given by:
\begin{align}
\tau\dot{s_i} &= \left( r_i - s_i/2 \right) \frac{S_\mathrm{max} - s_i}{S_\mathrm{max}} \\
\tau\dot{r_i} &= x_0 F\left( I - \sum_{j=1}^N g_{ij} s_j \right) \tau - r_i 
\end{align}
for $i = 1, 2, 3$, with 
\begin{equation}
    F(x) = \exp(-\epsilon/x) [\max(0, x)^\alpha ]. 
\end{equation}

The matrix $g$ is constructed such that $g_{21} = g_{32} = g_{13} = g_1$, $g_{12} = g_{23} = g_{31} = g_2$ and $g_{11} = g_{22} = g_{33} = 0$. The parameters are $\tau = 50$, $\epsilon = 10^{-3}$, $I = 0.145$, $S_\mathrm{max} = 0.045$, $g_1 = 3.0$, $g_2 = 0.7$, $x_0 = 2.57 \times 10^{-3}$, and $\alpha = 0.564$.

The numerical integration is best done with a change of coordinates $z_i \equiv \log(S_\mathrm{max} - s_i)$, which reduces the numerical precision difficulties associated with the trajectory getting too close to the stable manifold of the fixed points. The initial condition was $(0.5, 0.2, 0.4, 0.9, 0.5, 0.6)$. 

% The heteroclinic cycle as a whole is an attractor of the system. But the trajectories on the cycle alternate between the three saddle fixed points. Each saddle has distinct properties (their position in phase space), and the duration near them is long, so each saddle is a metastable state.

\subsubsection{Attractor-merging crisis}
We have provided an example of attractor-merging crisis for the Duffing oscillator, whose dynamics is described by \cite{ishii1986breakdown}: 
\begin{align}
    \dot{x} &= v \\ 
    \dot{v} &= -ax^3 + bx -dv + f \cos(\omega t), 
\end{align}

with parameters $a = 100$, $b = 10$, $d = 1.0$, $f = 0.852$, $\omega = 3.5$ and initial condition $(x,v) = (0.11, 0.11)$.

\subsubsection{Type-I intermittency}
The system used to obtain the type I intermittency is the Lorenz63 model \cite{lorenz1963deterministic}. The equations are given by 
\begin{align}
    \dot{x} &= \sigma(y - x) \\ 
    \dot{y} &= x(\rho - z) - y \\ 
    \dot{z} &= xz - \beta z. 
\end{align}

The parameters used in the figure are $\sigma = 10$, $\beta = 8/3$ and $\rho = 166.1$, based on \cite{yorke1979metastable}. The parameter used to obtain a stable limit cycle (before the saddle-node bifurcation) was $\rho = 166.06$. The initial condition was $(0.1, 0.1, 0.1)$.

% The trajectories alternate between the ghost of the limit cycle (where they have almost periodic behavior) and between chaotic behavior. In the example shown in Fig. \ref{fig:observationsmetastability}(c), the periodic regime lasts for a long time, compared to its own period, so it is a metastable state. The chaotic regime cannot be considered metastable in the figure, since its duration is too short. For other parameter choices, ggthe chaotic phase can be much longer, and then it would be characterized as metastable. 

\subsubsection{Chaotic saddle}
The chaotic saddle shown in the Fig. 4E-E\ssupprime occurs in the Ikeda system \cite{alligood1997book}, which has discretized time:
\begin{align}
    x_{n+1} &= a+b(x_n \cos(t_n) - y \sin(t_n)) \\ 
    y_{n+1} &= b(x_n \sin(t_n) + y_n \cos(t_n)),  
\end{align}
with 
\begin{equation}
    t_n = c - \frac{d}{1 + x_n^2 + y_n^2}.
\end{equation}
The parameters used for the chaotic saddle were $b = 0.9$, $c = 0.4$, $d = 6.0$, $a = 1.003$. The saddle exists due to a boundary crisis, which occurs for a slightly smaller $a$. For reference, the value of $a$ used to obtain a chaotic attractor was $a = 0.997$. The initial condition was $(2.97, 4.15)$. This is the only mechanism in the figure where the initial condition is important: to reproduce the behavior, the trajectories need to be initialized near the chaotic saddle.

\subsection{Brief notion of almost-invariant sets} \label{sec:brief-notion}
Following Ref. \cite{froyland2005statistically}, we provide a brief notion of almost-invariant sets for a mapping $T : X \to X$, with $X$ being the state space, but it can also be formulated for flows \cite{froyland2009almost}. To formalize the notion of almost-invariant sets, we need to first define a natural measure on any set $A$. This corresponds to the fraction of time that typical trajectories, starting in $x \in X$, spend on that set along their evolution:
\begin{equation}
    \mu (A) = \lim_{N\to\infty} \frac{\# \{i \in [0, N-1] : T^i x \in A \} }{N}
\end{equation}

The probability of trajectories, once inside $A$, to remain in $A$, is computed as 
\begin{align}
    \rho(A) = \mathrm{Prob}(Tx \in A | x \in A) = \frac{\mu( T^{-1} A \cap A)}{\mu(A)}.
\end{align}

A set $A$ is then said to be almost-invariant if $\rho(A)$ is high but not $1$, meaning that trajectories, once inside $R$, have a high probability of remaining in $R$. Suppose that the state space can be divided into disjointed almost-invariant sets $\{R_1, R_2, ..., R_N\}$. One can obtain maximally almost-invariant sets by maximizing the mean probability $\frac{1}{N}\sum_{i=1}^{N} \rho(A_i)$ \cite{froyland2005statistically, froyland2003detecting}.

% \subsubsection{Bursting}
% The bursting neuron shown in the figure is due to Hindmarsh and Rose \cite{hindmarshmodel1984}. The equations are 
% \begin{align}
%     \dot{V} &= y - aV^3 + bV^2 -z +I \\ 
%     \dot{y} &= c - dV^2 - y \\ 
%     \dot{z} &= r[s(V - x_r) -z]. 
% \end{align}

% The parameters used were $a=1$, $b=3$, $c=1$, $d=5$, $x_r=-8/5$, $s=4$, $r=0.001$, $I=2.0$, with initial conditions $(-1, 0, 0)$.

% This final example in Fig. \ref{fig:observationsmetastability}(e) shows a bursting behavior that is stable: the trajectory tends to a limit cycle composed of the green and purple regions. We separate these regions because each has a distinct property: the green corresponds to spikes (circling around in phase space), the purple to quiescence (converging a stable node). Because they have distinct properties, they can be considered as different states. We defend that even here the at least the green state is metastable, since is lasts several cycles. But we concede that this is a sort of a pathological case.

% \bibliographystyle{apalike}
%\bibliographystyle{unsrt}
% \bibliography{references}


\end{document}